\newif\ifPDFLaTeX
\expandafter\ifx\csname pdfoutput\endcsname\relax\else\PDFLaTeXtrue\fi
\documentclass[%
  aps,%
  prd,%
  showpacs,%
  preprintnumbers,%
  amsmath,%
  amssymb%
]{revtex4}
\usepackage{graphicx}
\usepackage{thophys}
\allowdisplaybreaks
\newcommand{\GeV}{\text{GeV}}
\newcommand{\TeV}{\text{TeV}}
\newcommand{\fb}{\text{fb}}
\newcommand{\ii}{\mathrm{i}}

\let\Re\relax
\DeclareMathOperator{\Re}{Re}

\begin{document}

\preprint{WUE--ITP--2003--020, TTP 03--25, hep-ph/0406098}

\title{Testing the Noncommutative Standard Model at a Future Photon Collider}

\author{Thorsten Ohl}
\email{ohl@physik.uni-wuerzburg.de}
\affiliation{%
  Institut f\"ur Theoretische~Physik und Astrophysik,
  Universit\"at~W\"urzburg, D--97074~W\"urzburg, Germany}

\author{J\"urgen Reuter}
\email{reuter@particle.uni-karlsruhe.de}
\affiliation{%
  Institut f\"ur Theoretische~Teilchenphysik,
  Universit\"at~Karlsruhe, D--76128~Karlsruhe, Germany}

\date{\today}

\begin{abstract}
  Extensions of the Standard Model of elementary particle physics to
  noncommutative geometries have been proposed as a low energy limit
  of string
  models. Independent of this motivation, one may consider such a model
  as an effective field theory with higher-dimensional operators
  containing an antisymmetric rank-two background field. We study the
  signals of such a Noncommutative Standard Model~(NCSM) and analyze
  the discovery potential of a future photon collider, considering
  angular distributions in fermion pair production.
\end{abstract}
\pacs{11.10.Nx, 11.30.Cp, 11.80.Cr, 13.88.+e, 13.90.+i}
\maketitle

%%%%%%%%%%%%%%%%%%%%%%%%%%%%%%%%%%%%%%%%%%%%%%%%%%%%%%%%%%%%%%%%%%%%%%%%%%%

\section{Introduction}
\label{sec:intro}

The idea of noncommutative space-time coordinates is more than half a
century old.  However, interest in noncommutative~(NC) theories
has been growing dramatically in recent years due to the observation that
open string theories with constant antisymmetric rank-two tensor background
fields in the limit of vanishing string tension $\alpha' \to 0$ can be
interpreted as Yang-Mills theories living on a NC
manifold~\cite{SeiWit}. Independent of this motivation provided by string
theory, Noncommutative Quantum Field Theory~(NCQFT) in itself provides
an interesting approach to introducing a fundamental length scale and
consequently cutting off short-distance contributions in a way that is
consistent with the symmetries of a given model.

Although there still remain open questions regarding the definition
and consistency of perturbative NCQFTs (e.\,g.~UV/IR
mixing~\cite{UVIR} and unitarity~\cite{Unitarity}) one can study a
particular NC structure and its phenomenological
consequences. This will be the scope of the present paper.

Recently, there has been a lot of activity in model buildung, trying
to construct an Effective Field Theory~(EFT) which is defined on a
NC spacetime with a canonical structure 
\begin{equation}
\label{eq:theta}
  [\hat{x}^\mu, \hat{x}^\nu]
    = \ii \theta^{\mu\nu}
    = \ii \frac{1}{\Lambda_{\text{NC}}^2} C^{\mu\nu}
\end{equation}
and has---ignoring potential violations of the decoupling theorem
engendered by UV/IR mixing in higher orders of perturbation
theory---the Standard Model~(SM) as low energy limit
for~$\sqrt{s}\ll\Lambda_{\text{NC}}$.  While it is reasonably
straightforward to construct a NC version of QED and there have been
several studies of the phenomenological consequences of
NCQED~\cite{Hinchliffe,HewettRizzo}, it is impossible to
comprise the whole~SM without additional constructions.  The key
issue here is the realization of gauge invariance on NC
spaces~\cite{SeiWit,Wess:pr}.  Early attempts suffered from the fact that only
certain gauge groups could be realized (in particular $\mathrm{U}(N)$,
but not~$\textrm{SU}(N)$~\cite{Armoni}) and from the quantization of
$\mathrm{U}(1)$ charges~\cite{Hayakawa:1999yt}.
The latter is caused by the Ward identity 
for the coupling of gauge fields to matter, which forces the triple
gauge boson coupling to be identical to the coupling of \emph{each}
particle to the gauge bosons.  Consequently, all particles must carry
the \emph{same} charge, which is incompatible with the hypercharge 
assignments in the~SM (see however~\cite{Sheikh-Jabbari} for a clever
symmetry breaking mechanism realizing the correct hypercharges and
allowing the construction of a noncommutative extension of the full SM
at the price of introducing additional heavy gauge bosons).

A general way to overcome the aforementioned problems is provided by
the Seiberg-Witten Map~(SWM)~\cite{SeiWit}. It is an asymptotic
expansion in the noncommutativity~$\theta^{\mu\nu}$ which relates the
fields on the NC spaces to the fields on a commutative
space.  $\mathrm{SU}(N)$ gauge groups and arbitrary $\mathrm{U}(1)$
charges can be realized by going from Lie algebras to their enveloping
algebras. The additional degrees of freedom introduced in this way can
be eliminated by the freedom in the SWM~\cite{Wess:pr}.  In this
approach, the r\^ole of the triple gauge boson couplings in the Ward
identities is taken over by new contact terms and the problem with
charge quantization does not appear.  A class of realistic models
based on the SWM and including the
full SM has been proposed soon after the introduction of the
SWM~\cite{NCSM}.  In the following, we will use the term
Noncommutative Standard Model~(NCSM) for this class of models.
Our goal in the current paper is to give a new
example of a search for signals of NC structures in the NCSM
model and also to present the methods needed for their calculation.

The organization of the current paper is as follows: in
section~\ref{sec:ncsm} we give a brief introduction to the NCSM.
In section~\ref{sec:neffsign} we
demonstrate which new effects will appear in the NCSM and propose
fermion pair production at a future photon collider as
an example for a process where
to search for signals of NC theories. The formalism used in our
analysis---helicity amplitudes with antisymmetric rank-two tensor
fields---is presented in section~\ref{sec:helamp}. We then analyze the
angular distribution in~$\gamma\gamma \to f\bar f$. In
section~\ref{sec:ward} we discuss consistency checks for our
calculation. In section~\ref{sec:xsect} we present our numerical
results before concluding and giving a short outlook to some further
possibilities in section~\ref{sec:concl}. We add an extensive appendix
with our conventions and details of the formalism used as well as a
list of the Feynman rules, which will serve as a reference for future
work~\cite{thojrprog}.

%%%%%%%%%%%%%%%%%%%%%%%%%%%%%%%%%%%%%%%%%%%%%%%%%%%%%%%%%%%%%%%%%%%%%%%%

\section{The Noncommutative Standard Model}
\label{sec:ncsm}

The NC structure of spacetime is associated with
a scale $\Lambda_{\text{NC}}$
\begin{equation}
\tag{\ref{eq:theta}'}
  [\hat{x}^\mu, \hat{x}^\nu]
    = \ii \theta^{\mu\nu}
    = \ii \frac{1}{\Lambda_{\text{NC}}^2} C^{\mu\nu}
    = \ii \frac{1}{\Lambda_{\text{NC}}^2} 
            \begin{pmatrix}
                0 & - E^1 & - E^2 & - E^3 \\ 
              E^1 & 0     & - B^3 &   B^2 \\
              E^2 &   B^3 & 0     & - B^1 \\
              E^3 & - B^2 &   B^1 & 0
            \end{pmatrix}
\end{equation}
and the noncommutativity~$\theta^{\mu\nu}$ is a real antisymmetric
matrix, assumed here to be constant, which can be understood as a
spurion breaking Lorentz invariance.  The dimensionless ``electric''
and ``magnetic'' parameters~$\vec E$ and~$\vec B$ have been introduced
in~(\ref{eq:theta}') for future convenience.  The microscopic origin
of the spurion~$\theta^{\mu\nu}$ is irrelevant as long as we are
merely studying the EFT, where 
it appears as a coefficient in front of operators of dimension six or
higher.

Functions on NC manifolds are realized by functions on a
commutative manifold, when their pointwise product is replaced by the
Moyal-Weyl $\star$-product:
\begin{equation}
  \left( f \star g \right) (x) = \left. \exp \left[ \frac{\ii}{2}
    \theta^{\mu\nu}
        \frac{\partial}{\partial x^\mu}
        \frac{\partial}{\partial y^\nu} \right] f(x) g(y) \right|_{y\to x}
   = f(x) g(x) + \dfrac{\ii}{2} \theta^{\mu\nu}
                     (\partial_\mu f(x)) (\partial_\nu g(x))
               + \mathcal{O}(\theta^2)\,.
\end{equation}
%\begin{multline}
%  \left( f \star g \right) (x) = \left. \exp \left[ \frac{\ii}{2}
%    \theta^{\mu\nu}
%        \frac{\partial}{\partial x^\mu}
%        \frac{\partial}{\partial y^\nu} \right] f(x) g(y) \right|_{y\to x} \\
%   = f(x) g(x) + \dfrac{\ii}{2} \theta^{\mu\nu}
%                     (\partial_\mu f(x)) (\partial_\nu g(x))
%               + \mathcal{O}(\theta^2)\,.
%\end{multline}
Gauge theories on a NC manifold can then be constructed with the help of the
SWM, which expresses the NC matter and gauge fields~$\hat\Psi$
and~${\hat A}_\mu$ as functions of commutative matter and gauge
fields~$\Psi$ and~$A_\mu$ so that the NC gauge
transformations~$\hat\Psi\to\hat\Psi'$ and~$\hat A\to\hat A'$ are
realized by the commutative gauge transformations~$\Psi\to\Psi'$
and~$A\to A'$
\begin{subequations}
\begin{align}
  \hat A'(A) &= \hat A (A') \\
  \hat\Psi' (\Psi, A) &= \hat\Psi (\Psi', A')\,.
\end{align}
\end{subequations}
To lowest order in~$\theta^{\mu\nu}$, the SWMs are
\begin{subequations}
  \begin{align}
    \hat{\Psi} &=\; \Psi + \dfrac{1}{2} \theta^{\mu\nu} A_\nu
    \partial_\mu \Psi + \dfrac{\ii}{8} \theta^{\mu\nu} \lbrack A_\mu ,
    A_\nu \rbrack \Psi + \mathcal{O}(\theta^2) \\ 
    \hat{A}_\lambda &=\; A_\lambda + \dfrac{1}{4} \theta^{\mu\nu} \left\{
    A_\nu , \partial_\mu A_\lambda \right\} + \dfrac{1}{4}
    \theta^{\mu\nu} \left\{ F_{\mu\lambda} , A_\nu \right\} +
    \mathcal{O}(\theta^2)\,.
  \end{align}
\end{subequations}
Equipped with this machinery, the construction of the NCSM is now
straightforward: one has to replace each field by the corresponding
SWM and all products by $\star$-products. This leads to
the Lagrangian given in~\cite{NCSM} from which the Feynman
rules can be derived. The
Feynman rules needed in the present paper are collected in
appendix~\ref{app:feyn}.

The contributions of the higher-dimensional operators are suppressed
by the ratios~$\Lambda^2_{EW}/\Lambda^2_{\text{NC}}$
and~$s/\Lambda^2_{\text{NC}}$ of the electroweak scale, the NC scale
and the CMS energy of the process. There have already been several papers
exploring the constraints from past and present-day experiments (at a
moment mainly from the non-observation of the Lorentz-violating $Z$
decays $Z\to \gamma\gamma, gg$ at LEP, as well as from astrophysics
\cite{Raffelt,raredec,TGC,anatho,ncbounds,PDG}). The bound on
$\Lambda_{\text{NC}}$ is still surprisingly low, of the order of $100
- 200 \, \GeV$. In a low lying string scenario one could expect values
for such a scale as low as $\Lambda_{\text{NC}} \gtrsim 1 \,\TeV$. 

%%%%%%%%%%%%%%%%%%%%%%%%%%%%%%%%%%%%%%%%%%%%%%%%%%%%%%%%%%%%%%%%%%%%%%%%

\section{New effects and signals}
\label{sec:neffsign}

Due to the presence of the higher-dimensional operators, there will be
deviations of decay rates and production cross sections from the SM
predictions. Existing SM vertices receive corrections with new Lorentz
structures and there are new vertices coupling more than one gauge
boson to matter fields.  The latter are required by the former in
order to retain gauge invariance.
In addition, there can also be new gauge boson interactions not allowed
in the SM, most prominently triple neutral gauge boson vertices such
as~$\gamma\gamma\gamma$, $Z\gamma\gamma$, $ZZZ$, $Zgg$, $\gamma
gg$~\cite{TGC}.

In general, taking the effects of the noncommutativity into account,
amplitudes for physical processes are asymptotic expansions
in~$\theta^{\mu\nu}$. Squared matrix elements
at~$\mathcal{O}(\theta^2)$ are (the subscripts 1 and 2 denote
the order in~$\theta$)
\begin{equation}
  |A|^2
     = |A^{\text{SM}}|^2
     + (A^{\text{SM}})^* A^{\text{NC}}_1
     + (A^{\text{NC}}_1)^* A^{\text{SM}}
     + |A^{\text{NC}}_1|^2 + (A^{\text{SM}})^* A^{\text{NC}}_2
     + (A^{\text{NC}}_2)^* A^{\text{SM}}.
\end{equation}
For processes forbidden in the SM, only the term~$|A^{\text{NC}}_1|^2$
contributes at this order. If there is a nonvanishing SM amplitude,
all interference terms have to be taken into account. Consequently,
the term with the first order NC amplitude squared has to be
dropped as long as we do not know the second order NC amplitudes,
which depend on the second order terms in the SWM for the
fields. Unfortunately, the SWMs for the NCSM are not yet known
beyond first order. Therefore we concentrate in the present paper on
the first order interference terms.

In the following, we will study fermion pair production $\gamma\gamma \to
f\bar f$ at a future photon collider. Such a machine has been proposed
for a future linear $e^+ e^-$-collider with energies up to $1
\,\TeV$ \cite{TDR}. Highly energetic photons are produced with
the help of Compton backscattering of laser photons off the LINAC
electron beam. Thanks to the Compton scattering cross section, the
photons can be delivered with a high degree of polarization. Indeed
polarization will be available from day one, because it is required
for obtaining a photon spectrum concentrated at high
energies~\cite{Ginzburg/Telnov/etal:gamma-collider}. This will be
crucial for our considerations.

The center of mass energy is planned to be in the range of several
hundreds of~$\GeV$.  We will assume massless fermions in our
calculation and postpone the discussion of the specific features of
top-quark production and decays to a future publication.
Theoretically, the approximation of massless fermions
together with the fact of having polarized photons suggests to use the
very elegant formalism of helicity amplitudes~\cite{BG,GaWu} for
evaluating the cross section.

Unfortunately, the realization of a photon collider still lies many
years in the future and we should consider other collider projects
that will provide data much sooner.  Experiments at the LHC will
search for signals of NC theories before the decade is out.
They are sensitive to different combinations
of~$\vec E$ and~$\vec B$~\cite{anatho} and will therefore be
complementary to experiments at a photon collider.

We shall see below that polarization is extremely helpful for
constructing sensitive observables.  As already mentioned, a high
degree of polarization works in favor of high luminosity at the photon
collider, while there is competition between
polarization and luminosity at the LHC.  Therefore the degree of
polarization at the
LHC should be expected to be much smaller compared to the photon
collider.  In principle, the production of two leptons at a photon
collider is less plagued by background contamination than vector boson
production at
the LHC, but a more detailled investigation is required for a
quantitative comparison~\cite{anatho}.

The $e^+e^-$ linear collider in the $e^+e^-$ mode will necessarily be
commissioned before a photon collider.  In this case, a high degree of
polarization will be part of the experimental program and will support
the search for signals of NC theories~\cite{thojrprog}.  Again the
experiments will be complementary, because processes with polarized
photons and processes with polarized fermions depend differently on
the parameters~$\vec E$ and~$\vec B$~\cite{thojrprog}.

%%%%%%%%%%%%%%%%%%%%%%%%%%%%%%%%%%%%%%%%%%%%%%%%%%%%%%%%%%%%%%%%%%%%%%%%

\section{Helicity amplitudes with noncommutativity}
\label{sec:helamp}

The major addition to the established helicity amplitude
machinery~\cite{BG,GaWu} required by our application is the spinor
representation of antisymmetric rank-two tensor fields in order to
incorporate
the noncommutativity into the spinor products. With the help of the Schouten
identity (\ref{eq:schouten}) the noncommutativity can be converted in a very
elegant way to a spinor expression containing only the spinor metric and a
symmetric rank-2 spinor (for a textbook presentation cf.~\cite{Penrose:book}):
\begin{equation}
  \theta_{A\dot{A},B\dot B}
     = \theta^{\mu\nu}
        \bar{\sigma}_{\mu,A\dot A}
        \bar{\sigma}_{\nu,B\dot B} 
     = \phi_{AB} \epsilon_{\dot A\dot B}
        + \bar{\phi}_{\dot A\dot B} \epsilon_{AB}
\end{equation}
with~$(\phi_{AB})^* = \bar{\phi}_{\dot A\dot B}$.  Then~$\phi_{AB} = \frac12
\theta_{A\dot C,B}^{\hphantom{A\dot C,B}\dot C}$ 
is symmetric with three independent complex components
\begin{subequations}
\begin{align}
  \phi_{11} & = - E_- - \ii B_- \\
  \phi_{12} & = E_3 + \ii B_3 =\phi_{21} \\
  \phi_{22} & = E_+ + \ii B_+\,,
\end{align}
\end{subequations}
using the parameterization~(\ref{eq:theta}'), with $E_\pm = E^1 \pm \ii
E^2$, $B_\pm = B^1 \pm \ii B^2$.
With the help of this expression, all amplitudes can be expressed as contractions
of Weyl-van der Waerden spinors with~$\phi$ and among themselves.

To convert contractions of the noncommutativity with two vectors (momenta
and polarization vectors)
\begin{equation}
  \label{pqvector}
  (V_1\theta V_2) := V_{1,\mu} \theta^{\mu\nu} V_{2,\mu}
     = V_1^0 (\vec E \cdot \vec V_2) - V_2^0 (\vec E \cdot \vec V_1)
       - \vec B \cdot (\vec V_1 \times \vec V_2)
\end{equation}
into spinor expressions 
\begin{equation*}
  (V_1 \theta V_2) = \frac{1}{2} \Re \left[ \Braket{v_1 v_2}^*
  \Braket{v_1 \phi v_2} \right] 
\end{equation*}
we introduce the symmetric spinor products
\begin{equation}
  \Braket{p \phi q} = p^A \phi_{AB} \,q^B = \Braket{q \phi p} , \qquad
  \Braket{p \phi q}^* =
     p^{\dot A} \bar\phi_{\dot A\dot B} q^{\dot B}.
\end{equation}
As this is a non-standard spinor product we give an explicit
expression (in the conventions described in appendix~\ref{app:spiprod})
\begin{equation}
  \Braket{p\phi q}
     = \epsilon^{BA} p_A \phi_{BC} \epsilon^{CD} q_D 
     = - p_A \epsilon^{AB} \phi_{BC}
          \epsilon^{CD} q_D 
     = \phi_{11} p_2q_2 + \phi_{22} p_1q_1 - \phi_{12} (p_1q_2 + p_2q_1).
\end{equation}
%\begin{multline}
%  \Braket{p\phi q}
%     = \epsilon^{BA} p_A \phi_{BC} \epsilon^{CD} q_D 
%     = - p_A \epsilon^{AB} \phi_{BC}
%          \epsilon^{CD} q_D \\
%     = \phi_{11} p_2q_2 + \phi_{22} p_1q_1 - \phi_{12} (p_1q_2 + p_2q_1).
%\end{multline}

We study the processes $\gamma(k_1)\gamma(k_2)\to f(p_1) \bar f(p_2)$.
In the SM, there are $t$- and $u$-channel exchange diagrams
\begin{equation}
  A^{\text{SM}}_t =
    \parbox{30mm}{\includegraphics[width=30mm]{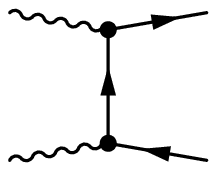}} \qquad
  A^{\text{SM}}_u =
    \parbox{30mm}{\includegraphics[width=30mm]{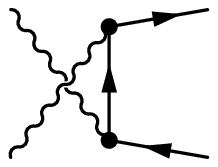}}\,.
\end{equation}
Due to helicity conservation, the only nonvanishing combinations
for massless fermions in the final state are~$(\pm,\mp) \to (\pm,\mp)$
and~$(\pm,\mp) \to (\mp,\pm)$. Since we are interested in the
interference of the NCSM with the SM amplitude, we do not have to
calculate the NC amplitude for the other combinations.

There are two $O(\theta)$
contributions for each standard model diagram
(the Feynman rules are collected in appendix~\ref{app:feyn})
\begin{subequations}
\begin{align}
  A^{\text{NC}}_{t,1} &=
    \parbox{30mm}{\includegraphics[width=30mm]{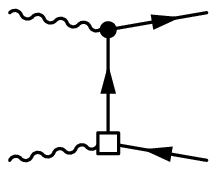}} \qquad
  A^{\text{NC}}_{t,2} =
    \parbox{30mm}{\includegraphics[width=30mm]{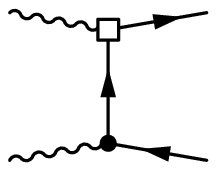}}\\
  A^{\text{NC}}_{u,1} &=
    \parbox{30mm}{\includegraphics[width=30mm]{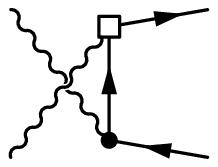}} \qquad
  A^{\text{NC}}_{u,2} =
    \parbox{30mm}{\includegraphics[width=30mm]{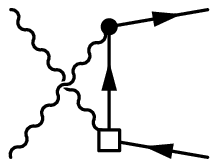}}\,.
\end{align}
The contact term is required by gauge invariance
\begin{equation}
  A^{\text{NC}}_c =
    \parbox{30mm}{\includegraphics[width=30mm]{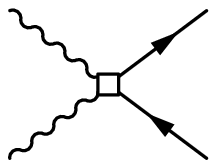}}\,,
\end{equation}
and there are two additional $s$-channel diagrams that depend on new coupling
constants~$K_{\gamma\gamma\gamma}$ and~$K_{Z\gamma\gamma}$
(matching the NCSM to the SM constrains the new triple
gauge boson couplings and the two couplings can not vanish
simultaneously~\cite{TGC})
\begin{equation}
\label{eq:NC-s-channel}
  A^{\text{NC}}_{s,\gamma} =
    \parbox{30mm}{\includegraphics[width=30mm]{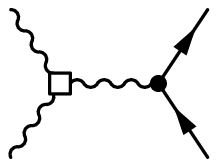}} \qquad
  A^{\text{NC}}_{s,Z} =
    \parbox{30mm}{\includegraphics[width=30mm]{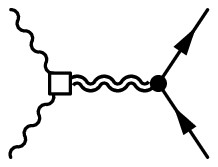}}\,.
\end{equation}
\end{subequations}
The explicit expressions for the helicity amplitudes for the
choice~$g_-=p_1$ and~$g_+=p_2$ of gauge spinors are
\begin{subequations}
\begin{align}
  A^{\text{NC},(+,-)}_{s,Z} + A^{\text{NC},(+,-)}_{s,\gamma}
    &= K_{\gamma Z}^{(+,-)} \biggl[
         c_1 \cdot \Braket{p_1k_1}^* \Braket{p_2 k_1}
       + c_2 \cdot \frac{\Braket{p_1 p_2}^* \Braket{p_2k_1}}{\Braket{p_2k_1}^*}
       \notag \\
    &\hphantom{= K_{\gamma Z}^{(+,-)} \biggl[}
       + c_3 \cdot \frac{\Braket{p_1 k_2}^* \Braket{p_2 p_1}}{\Braket{p_1k_2}}\biggr] \\  
  A^{\text{NC},(-,+)}_{s,Z} + A^{\text{NC},(-,+)}_{s,\gamma}
    &= K_{\gamma Z}^{(-,+)}
     \cdot c_1 \cdot \Braket{p_2k_1}^*\Braket{p_1k_1}
\end{align}
\end{subequations}
with the coefficients
\begin{subequations}
\begin{align}
  c_1 &=   2 (k_1 \theta k_2) (\varepsilon_1 \varepsilon_2)
         - s (\varepsilon_1 \theta \varepsilon_2) \\ 
  c_2 &= \sqrt{2} \bigl[
            (k_1 \theta \varepsilon_2) s - 2 (k_1 \theta k_2) (\varepsilon_2 k_1)
         \bigr] \\ 
  c_3 &= \sqrt{2} \bigl[
            (k_2 \theta \varepsilon_1) s + 2 (k_1 \theta k_2) (\varepsilon_1 k_2)
         \bigr]
\end{align}
\end{subequations}
and propagator factors
\begin{subequations}
\begin{align}
  C_{\gamma\gamma\gamma}
    &= \frac{2 e^2 Q_f \sin(2\theta_W) K_{\gamma\gamma\gamma}}{s} \\
  K_{\gamma Z}^{(+,-)}
    &= C_{\gamma\gamma\gamma}
     + \frac{4 g^2 s^4_W Q_f K_{Z\gamma\gamma}}{s -M_Z^2 + \ii M_Z \Gamma_Z}\\
  K_{\gamma Z}^{(-,+)}
    &= C_{\gamma\gamma\gamma}
     - \frac{4 g^2 s^2_W K_{Z\gamma\gamma}}{s -M_Z^2 + \ii M_Z \Gamma_Z}
       \left( T^3 - s_W^2 Q_f \right) \,.
\end{align}
\end{subequations}
The remaining diagrams contribute
\begin{subequations}
\begin{align}
  A^{(+,-)}_{t,1}
    &= \frac{- e^2 Q_f^2}{\sqrt{2} t}
       \frac{\Braket{p_1k_1} \Braket{p_2p_1} \Braket{p_1k_2}^*}{\Braket{p_1k_2}}
         \times\notag\\
    &\hphantom{=}\quad
           \biggl[ (\epsilon_1\theta p_1) \Braket{p_1 k_1}^*
                  - \sqrt{2}(k_1\theta p_1) \frac{\Braket{p_1p_2}^*}
                         {\Braket{p_2 k_1}^*} \biggr] \\ 
  A^{(-,+)}_{t,1}
    &= \frac{- e^2 Q_f^2}{t} 
       \frac{\Braket{p_1k_1} \Braket{p_2k_2}^* \Braket{p_1k_1}}{\Braket{p_1k_2}}
       (k_1\theta p_1) \\
  A^{(+,-)}_{u,1}
    &= \frac{- e^2 Q_f^2}{\sqrt{2} u}
         \frac{\Braket{k_1p_2} \Braket{p_1k_2}^*}{\Braket{p_2k_1}^*}
         \biggl[ (\epsilon_2\theta p_1) \Braket{k_2p_1} \Braket{p_1p_2}^*
                   + \sqrt{2} (k_2\theta p_1) \Braket{k_2p_2}^* \biggr] \\
  A^{(-,+)}_{u,1}
    &= 0 \\
  A^{(+,-)}_{t,2}
    &= \frac{- e^2 Q_f^2}{\sqrt{2}t}
       \frac{\Braket{p_1 k_2}^* \Braket{p_1p_2}^* \Braket{p_1 k_1}}
            {\Braket{p_2 k_1}^*}
         \biggl[ (\epsilon_2\theta p_2) \Braket{k_2 p_2}
                   - \sqrt{2} (k_2 \theta p_2)
                       \frac{\Braket{p_1p_2}}{\Braket{p_1 k_2}} \biggr] \\ 
  A^{(-,+)}_{t,2}
    &= \frac{-e^2 Q_f^2}{t}
       \frac{\Braket{k_1p_1}\Braket{p_2k_2}^*\Braket{k_2p_2}^*}{\Braket{p_2k_1}^*}
       (k_2\theta p_2) \\
  A^{(+,-)}_{u,2}
    &= \frac{- e^2Q_f^2}{\sqrt{2}u}
       \frac{\Braket{k_1p_2}\Braket{p_1k_2}^*}{\Braket{p_1k_2}}
         \biggl[ (\epsilon_1\theta p_2)\Braket{p_1p_2}\Braket{p_2k_1}^*
                   - \sqrt{2}(k_1\theta p_2)\Braket{p_1k_1}
         \biggr] \\
  A^{(-,+)}_{u,2}
    &= 0 \\
  A^{(+,-)}_c
    &= \frac{e^2 Q_f^2}{2} \biggl[
          (\epsilon_1\theta\epsilon_2)
             \left(\Braket{p_1k_1}^* \Braket{p_2k_1} - \Braket{p_1 k_2}^* \Braket{p_2 k_2} \right)
            \notag \\
    &\hphantom{= \frac{e^2 Q_f^2}{2} \biggl[}
      + \sqrt{2} ((k_1-k_2)\theta\epsilon_1)
         \frac{\Braket{p_1k_2}^* \Braket{p_2p_1}}
              {\Braket{p_1k_2}}
            \notag \\
    &\hphantom{= \frac{e^2 Q_f^2}{2} \biggl[}
      - \sqrt{2} ((k_1-k_2)\theta\epsilon_2)
         \frac{\Braket{p_1p_2}^*\Braket{p_2k_1}}
              {\Braket{p_2 k_1}^*}
      \biggr] \\
  A^{(-,+)}_c
    &= \frac{e^2 Q_f^2}{2} (\epsilon_1 \theta \epsilon_2) \biggl[
  \Braket{p_2 k_1}^* \Braket{p_1 k_1} - \Braket{p_2 k_2}^* \Braket{p_1 k_2}
  \biggr]\,.
\end{align}
\end{subequations}
The amplitudes with the other combination of photon helicities can be
obtained by simply interchanging $k_1$ with $k_2$.  The well
known SM amplitudes have been reproduced in appendix~\ref{app:helamp}.

From the analytical form, one can deduce that the interference depends
only on the space-time non-commutativity $\vec{E}$, but not on the
space-space noncommutativity $\vec{B}$. In fact, a nonvanishing
interference appears only for non-zero $E^1$ or $E^2$ . Note that this
dependence on the components of $\theta^{\mu\nu}$ in the NCSM is
different from NCQED~\cite{HewettRizzo}.

\begin{figure}
  \begin{center}
    \includegraphics[width=0.9\columnwidth]{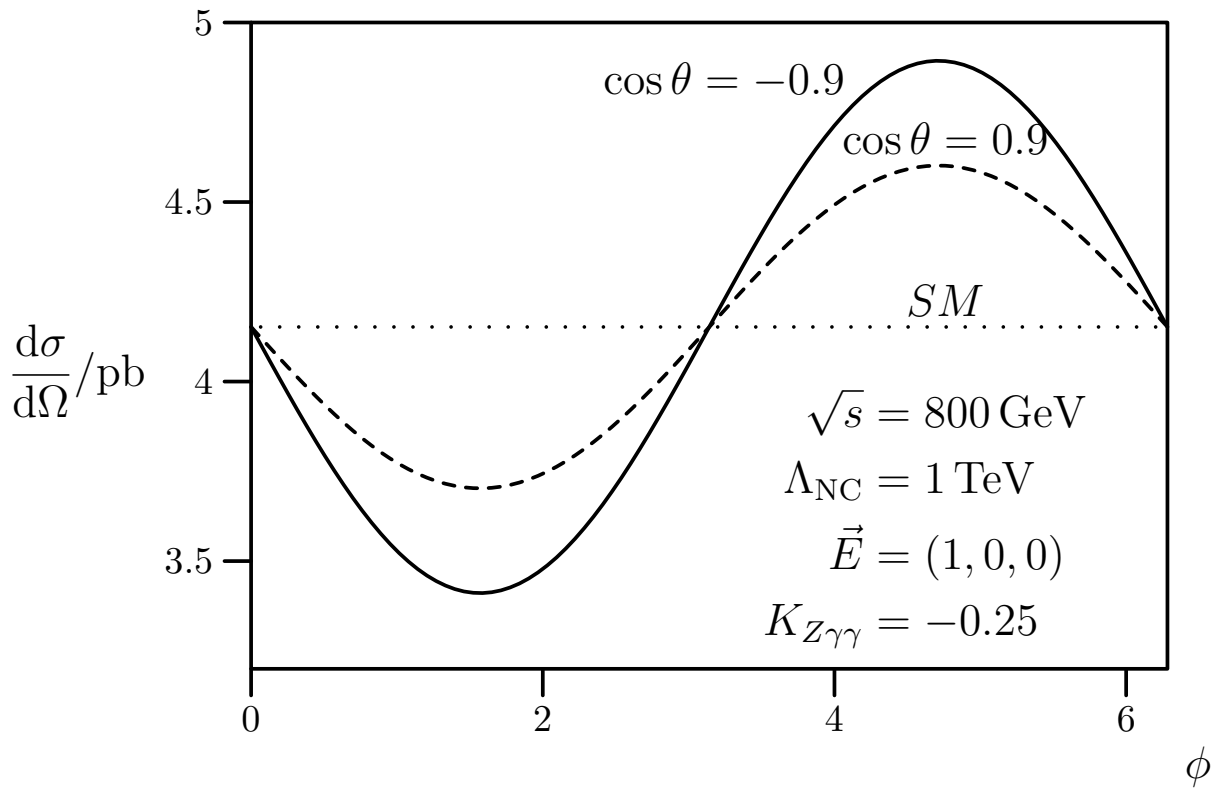}
  \end{center}
  \caption{\label{fig:800}%
    Dependence of the differential NCSM cross section on the azimuthal
    angle~$\phi$ for the time-like noncommutativity~$\vec E$
    perpendicular to the beam axis.}
\end{figure}
In figure~\ref{fig:800}, the differential cross section at a
center-of-mass energy of $800\,\GeV$ is plotted against the
azimuthal angle $\phi$. One sees that the integration over the
full solid angle will yield the same result for the interference as
for the SM cross section, because the squared matrix element is
proportional to~$\sin(\phi + \phi_0)$, where~$\phi_0$ is a phase which
depends on the spatial orientation of the noncommutativity.

\begin{figure}
  \begin{center}
    \includegraphics[width=0.9\columnwidth]{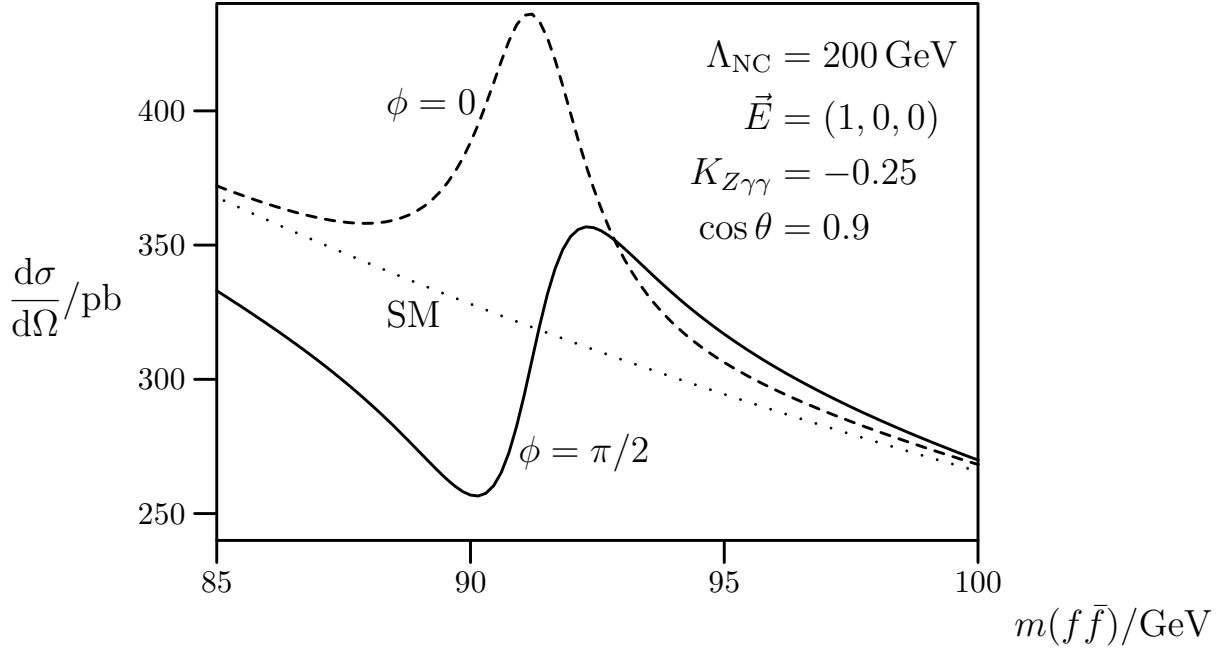}
  \end{center}
  \caption{\label{fig:zres}%
    The interference of the SM and $\mathcal{O}(\theta)$ NCSM amplitudes
    around the~$Z$ resonance is plotted for different values of
    the azimuthal angle~$\phi$.}
\end{figure}
In the same way as it is possible for a $Z$ in the NCSM to decay into
two photons violating the Yang-Landau theorem, it can be produced
resonantly in photon collisions. Figure~\ref{fig:zres} shows the
shapes of the interference contributions near the~$Z$ resonance
for different values of the azimuthal angle.  The effect for $\phi=0$
is only due to the imaginary part of the SM amplitude close to the
resonance and is therefore absent in figure~\ref{fig:800}, 
where~$\sqrt{s}\gg M_Z$.  Unfortunately, this resonance will hardly
be visible for $\Lambda_{\text{NC}}\gg M_Z$.

%%%%%%%%%%%%%%%%%%%%%%%%%%%%%%%%%%%%%%%%%%%%%%%%%%%%%%%%%%%%%%%%%%%%%%%

\section{Consistency Checks: Amplitudes and Ward Identity}
\label{sec:ward}

In order to control the numerical stability of a simulation as
well as to assure the correctness of our results, we have
performed several cross checks.
Our first check was to compare numerically the amplitude calculated
between two completely different formalisms: helicity amplitudes on the
one hand and Dirac spinors and polarization vectors on the other
hand (using an extension of the optimizing matrix element
compiler \texttt{O'Mega}~\cite{Omega} and event generator
\texttt{WHIZARD}~\cite{WHIZARD}). We found the results to agree within
numerical accuracy. 

Another way to check the resulting amplitude is to prove that the
Ward identity is fulfilled.
Writing the Dirac matrix strings for the amplitudes
\begin{subequations}
\begin{align}
  \ii A^t_{\mu_2\mu_1} &=
    \bar u(p_1) \biggl[
        \ii g\gamma_{\mu_2}
        \frac{\ii}{\fmslash{p}_1-\fmslash{k}_2}
        \ii g\Gamma_{\mu_1}(-k_1,p_2) \notag \\
      &\qquad\qquad\qquad \mbox{}
      + \ii g\Gamma_{\mu_2}(-k_2,-k_1+p_2)
        \frac{\ii}{\fmslash{p}_1-\fmslash{k}_2}
        \ii g\gamma_{\mu_1}
    \biggr] v(p_2) \\
  \ii A^u_{\mu_2\mu_1} &=
    \bar u(p_1) \biggl[
        \ii g\gamma_{\mu_1}
        \frac{\ii}{\fmslash{p}_1-\fmslash{k}_1}
        \ii g\Gamma_{\mu_2}(-k_2,p_2) \notag \\
      &\qquad\qquad\qquad \mbox{}
      + \ii g\Gamma_{\mu_1}(-k_1,-k_2+p_2)
        \frac{\ii}{\fmslash{p}_1-\fmslash{k}_1}
        \ii g\gamma_{\mu_2}
    \biggr] v(p_2) \\
  \ii A^c_{\mu_2\mu_1} &=
    \bar u(p_1) \ii g^2 H_{\mu_2\mu_1}(-k_2,-k_1) v(p_2)
\end{align}
\end{subequations}
and using (cf.~appendix~\ref{app:feyn})
\begin{subequations}
\begin{align}
  k^\mu \Gamma_\mu (k, p) &= 0 \\
  \varepsilon^\mu \Gamma_\mu (k, p) &=
    - \frac{\ii}{2} \left[
         (k\theta p) \fmslash{\varepsilon}
         - (k\theta\varepsilon) \fmslash{p}
         - (\varepsilon\theta p) \fmslash{k}
       \right] \\
  k_1^{\mu_1} \varepsilon_2^{\mu_2} H_{\mu_1\mu_2} (k_1, k_2) &=
    - \frac{\ii}{2} \left[
      \mbox{}
           (k_1\theta k_2) \fmslash{\varepsilon}_2
         - (\varepsilon_2\theta k_2) \fmslash{k}_1
         - (k_1\theta\varepsilon_2) \fmslash{k}_2
       \right] \notag \\
    &=   \varepsilon_2^{\mu_2} \Gamma_{\mu_2} (k_1, k_2)
     = - \varepsilon_2^{\mu_2} \Gamma_{\mu_2} (k_2, k_1)\,,
\end{align}
\end{subequations}
one sees analytically that the Ward identity is satisfied indeed:
\begin{multline}
  k_2^{\mu_2} \varepsilon_1^{\mu_1}
     (   A^t_{\mu_2\mu_1}
       + A^u_{\mu_2\mu_1}
       + A^c_{\mu_2\mu_1} ) = \\
      g^2 \bar u(p_1) \left[ \mbox{}
         \varepsilon_1^{\mu_1}\Gamma_{\mu_1}(-k_1,p_2)
       - \varepsilon_1^{\mu_1}\Gamma_{\mu_1}(-k_1,-k_2+p_2)
       + \varepsilon_1^{\mu_1}\Gamma_{\mu_1}(-k_1,-k_2)
    \right] v(p_2)  = 0.
\end{multline}
The gauge independence manifests itself in
the independence of the helicity amplitudes from the choice of the
gauge spinor. We have verified this independence for our results
within the numerical accuracy.

%%%%%%%%%%%%%%%%%%%%%%%%%%%%%%%%%%%%%%%%%%%%%%%%%%%%%%%%%%%%%%%%%%%%%%%%
\section{Results: Cross Section and Event Generation}
\label{sec:xsect}

To get a realistic cross section from the squared matrix element, one
has to fold the resulting cross section with the photon spectrum
produced in Compton scattering for the photon collider. For this
purpose, the program {\tt Circe 2.0} has been used \cite{Circe}, which
parameterizes the results of a microscopic simulation of the beam
dynamics~\cite{Telnov}.

\begin{figure}
  \begin{center}
    \includegraphics[width=0.9\columnwidth]{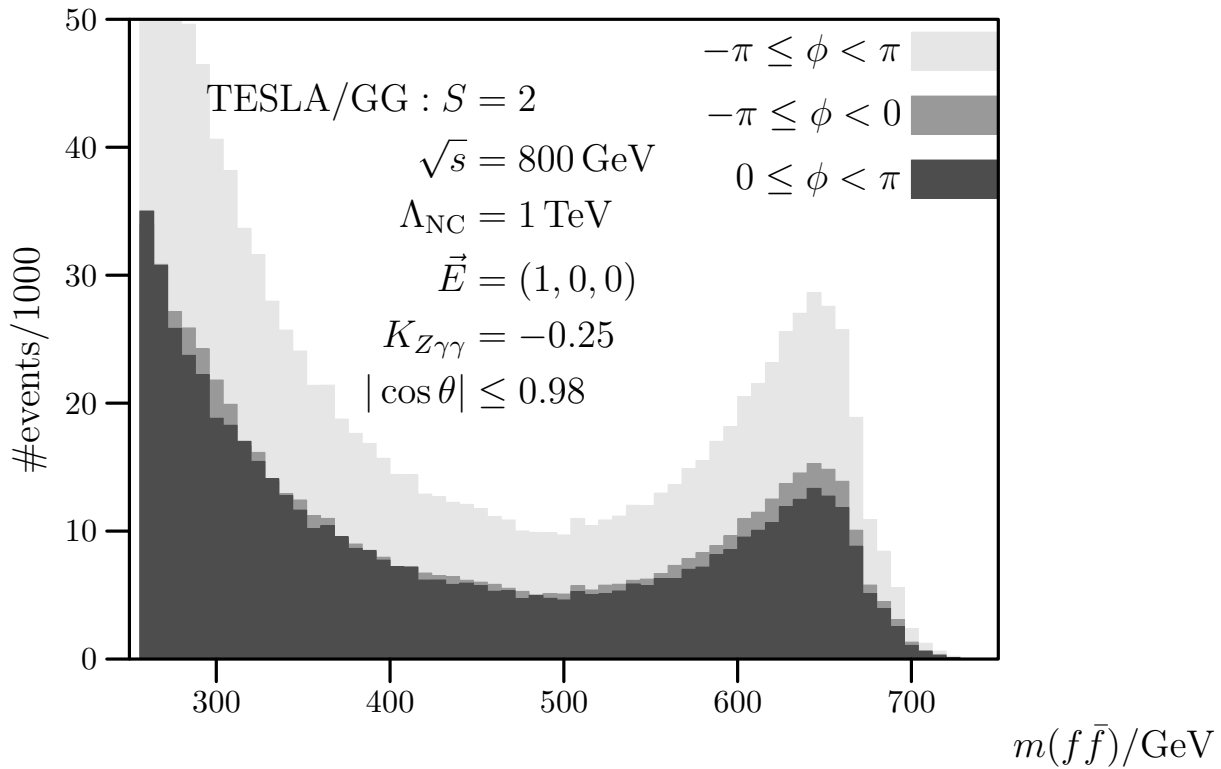}
  \end{center}
  \caption{\label{fig:s800fl}%
     Number of events per year in the two halfspheres $\phi<0$ and
     $\phi>0$ for $\sqrt{s}=800\,\GeV$.}
\end{figure}

As already mentioned in section~\ref{sec:neffsign}, the integrated
cross section
at order $\theta$ does not differ from the SM cross section since it
is linear in trigonometric functions of the azimuthal angle. However, in
the polarized cross section it is possible to scan over the final
state fermions to look for angular dependent deviations from the SM
prediction. 

In figure \ref{fig:s800fl} we show the number of binned events
over the invariant mass of the fermion pair, assuming integrated
luminosities of $400\,\fb^{-1}$ for~$200\,\GeV$, $1000\,\fb^{-1}$
for~$500\,\GeV$ and~$2000\,\fb^{-1}$ for~$1\,\TeV$.  These are
expected~\cite{TDR} for one year of running with~$30\%$ uptime.
For the $Z\gamma\gamma$ couplings we chose the central
value~$K_{Z\gamma\gamma} = - 0.25$, but the cross section and the
angular variation away from the $Z$ resonance do not depend very much
on this choice. A cut of about $11.5$~degrees around
each beam axis has been applied. 

Here one should notice one important point: A non-zero signal can only
be seen in the case that the helicities of the two photons are
different, so that they add up to a total helicity of two. Therefore, the
photon collider has to be run in the $d$-wave mode ($S=2$) with
different helicities. Originally, the photon collider was proposed to be
run in the $s$-wave mode ($S=0$) which is mandatory in the study of
the quantum numbers of scalar
or pseudoscalar particles like the Higgs. Fortunately, the photon
collider can be operated in each of the two modes and one can easily
switch between them.

\begin{figure}
  \begin{center}
    \includegraphics[width=0.9\columnwidth]{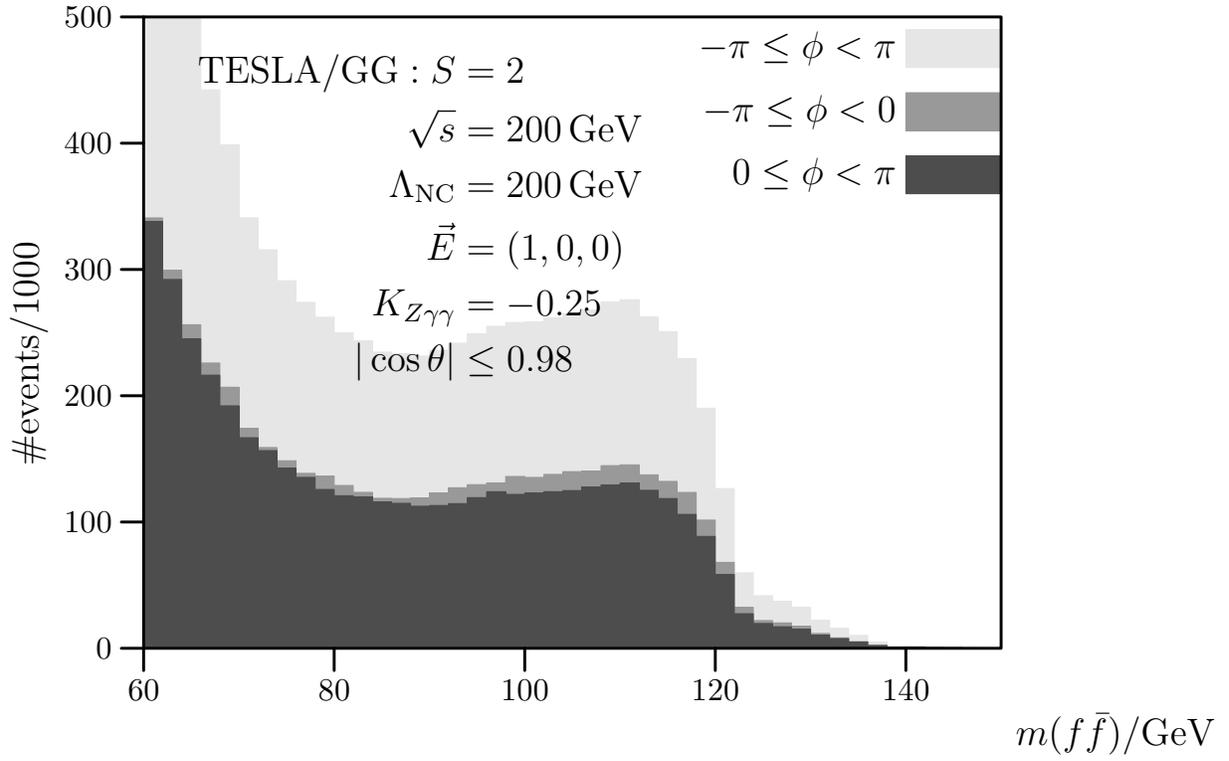}
  \end{center}
  \caption{\label{fig:s200}%
     Number of events per year in the two halfspheres $\phi<0$ and
     $\phi>0$ for $\sqrt{s}=200\,\GeV$.}
\end{figure}

\begin{figure}
  \begin{center}
    \includegraphics[width=0.9\columnwidth]{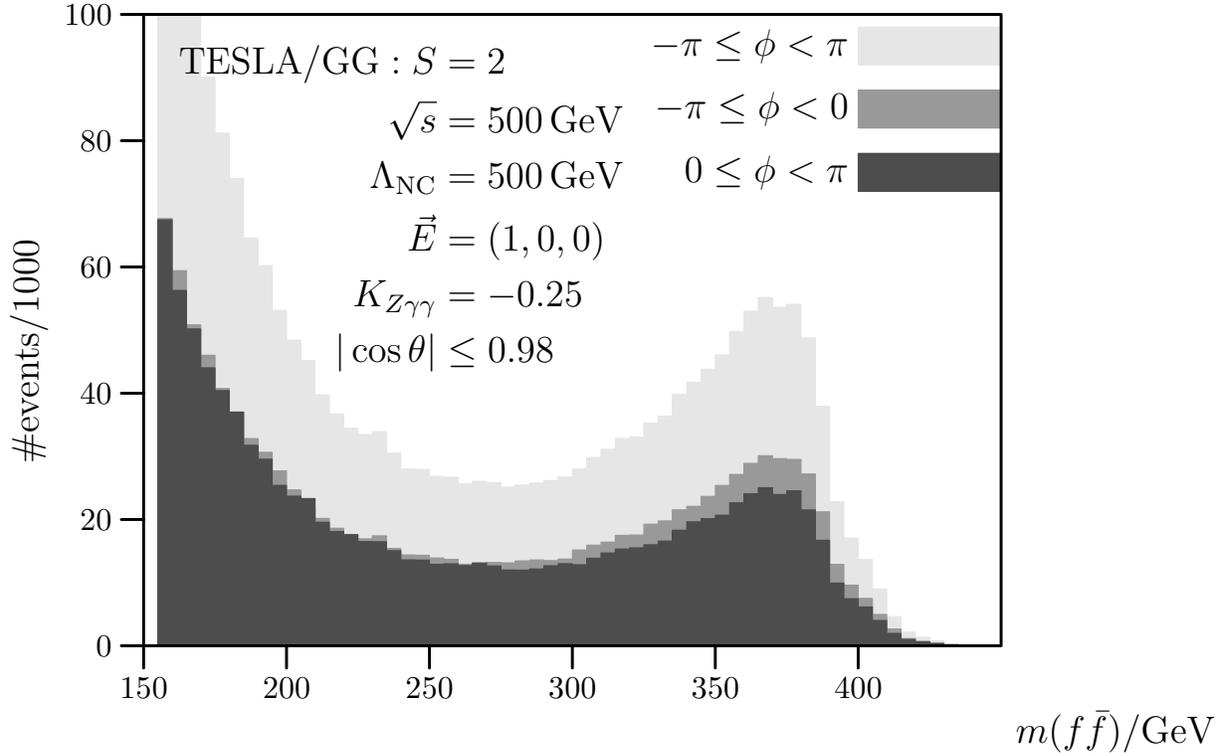}
  \end{center}
  \caption{\label{fig:s500}%
     Number of events per year in the two halfspheres $\phi<0$ and
     $\phi>0$ for $\sqrt{s}=500\,\GeV$.}
\end{figure}

In order to estimate the maximum effect we have assumed that the axis
of alignment for the detector relative to the noncommutativity is
known and have subdivided the full solid angle accordingly. In
practice, one first has to scan through the angular distributions of
the fermions in order to find deviations from the isotropic
distribution around the beam axis. Then, one can maximize the
anisotropy and in such a way fit the angular distribution to the data.
In this process one has to take into account the fact that, due to the
rotation of the earth and the rotation of the earth around the sun,
the direction of the noncommutativity relative to the collider and the
detector is almost certainly not constant.  Instead one must use a
coordinate system fixed in space.  Since the orientation of the
noncommutativity can therefore not be optimized for a maximal effect,
the results shown in figures~\ref{fig:s800fl}-\ref{fig:s500} are an
optimistic upper limit and should be expected to be diluted by a
factor of two in addition to the usual systematic uncertainties.

Figures~\ref{fig:s800fl}-\ref{fig:s500} show that a signal can be seen
easily if the scale~$\Lambda_{\text{NC}}$ is not far above the
CMS-mass energy~$\sqrt{s}$ of the linear collider, but generically the result
gets worse if the scale is lower. Our present calculation must not be
used for collider energies higher than the
scale~$\Lambda_{\text{NC}}$, since higher orders in~$\theta$ can only
be neglected if~$s/(\Lambda_{\text{NC}})^2 \lesssim 1$.

Other useful processes in the search for a signal of the~NCSM
are~$\gamma\gamma\to \gamma\gamma$, $\gamma\gamma\to\gamma Z$
and~$\gamma\gamma\to ZZ$. However, the dimension-six operators
consisting of three field strength tensors engendered by the kinetic
terms in the NCSM cannot contain four neutral electroweak gauge bosons,
because~$SU(2)$ has rank one. Therefore interferences with such
(loop-induced) SM amplitudes occur only at order~$\theta^2$.

%%%%%%%%%%%%%%%%%%%%%%%%%%%%%%%%%%%%%%%%%%%%%%%%%%%%%%%%%%%%%%%%%%
\section{Conclusions}
\label{sec:concl}

An extension of the Standard Model to noncommutative spacetime---which
arises in certain low-energy limits of string theories---offers a
variety of new phenomena.
Due to the presence of an antisymmetric rank-two spurion field
which breaks Lorentz invariance at a scale $\Lambda_{\text{NC}}$
the~SM is supplemented by a number of dimension-six
operators which result in deviations of decay rates, production cross
sections and other observables from their SM predictions. In this
paper we have focused on fermion pair production at a future photon
collider as an example for exploring the sensitivity of future
accelerator experiments to the free parameters~$\theta^{\mu\nu}$,
$K_{Z\gamma\gamma}$, etc.~of such models.

A generic signal for noncommutativity is the violation of angular
momentum conservation, which stems from the noncommutativity acting as
a static source of angular momentum and which leads to
violations of the isotropic distribution of final state particles
around the beam axis. 

Polarization is a helpful---if not mandatory---ingredient of
searches for signals of noncommutative theories. Therefore,
the high degree of polarization for the electron (and possibly also for
the positron) beam at the future linear collider facilitates
searches for the NCSM
directly at the lepton collider~\cite{thojrprog}. The methods presented
here will be used in the corresponding calculations.
Nevertheless, the very low background environment of photon collisions
provides a
good example for NCSM searches. In a conservative estimate, a photon collider
will be sensitive to scales of the order of $\Lambda_{\text{NC}} \sim
1\,\TeV$, but once enough data will be available, experimental
ingenuity will certainly push this limit upwards.

%%%%%%%%%%%%%%%%%%%%%%%%%%%%%%%%%%%%%%%%%%%%%%%%%%%%%%%%%%%%%%%%%%%%%%%%
\subsection*{Acknowledgments}

T.\,O. is supported by the Deutsche Forschungsgemeinschaft~(DFG), grant
RU\,311/1-1, and the Bundesministerium f\"ur Bildung und Forschung
Germany, grant 05HT1RDA/6.  J.\,R.~is supported by the DFG
Sonderforschungsbereich (SFB) ``Trans\-regio 9 -- Computergest\"utzte
Theoretische Teilchenphysik'' and the Graduier\-tenkolleg (GK)
``Hoch\-energiephysik und Teilchenastrophysik''.

We are grateful to Ana Alboteanu and Klaus M\"onig for valuable
discussions.

%%%%%%%%%%%%%%%%%%%%%%%%%%%%%%%%%%%%%%%%%%%%%%%%%%%%%%%%%%%%%%%%%%%%%%%%
\appendix
\section{Spinors and Spinor Products} 
\label{app:spiprod}
\subsection{General conventions}

Complex conjugation interchanges dotted and undotted indices:
$\bar{\xi}_{\dot{A}} = (\xi_{A})^*$, $\xi^{A} = (\bar{\xi}^{\dot{A}})^*$.
The spinor metric is $\epsilon^{AB} = \epsilon^{\dot{A}\dot{B}} =
\epsilon_{AB} = \epsilon_{\dot{A} \dot{B}}$ with
$\epsilon_{AB} = - \epsilon_{BA}$ and $\epsilon_{12} = +1$.
Our convention for lowering or raising of spinor indices is
\begin{align}
\label{eq:hebsenk}
\xi^{A} &=\; \epsilon^{AB} \xi_{B}, \quad
\bar{\xi}^{\dot{A}} = 
\epsilon^{\dot{A} \dot{B}} \bar{\xi}_{\dot{B}}, \notag \\  
\xi_{A} &=\; \xi^{B} \epsilon_{BA}, \quad 
\bar{\xi}_{\dot{A}} =
\bar{\xi}^{\dot{B}} \epsilon_{\dot{B} \dot{A}} .
\end{align}
The antisymmetric spinor product for commuting components
$\eta \xi = \eta_{A} \xi^{A} = \eta_1 \xi_2 - \eta_2 \xi_1$, 
$\bar{\eta} \bar{\xi} = \bar{\eta}_{\dot{A}} \bar{\xi}^{\dot{A}} =
(\eta_1 \xi_2 - \eta_2 \xi_1 )^* = \bar{\eta}_{\dot{1}} 
 \bar{\xi}_{\dot{2}} - \bar{\eta}_{\dot{2}} \bar{\xi}_{\dot{1}}$ and
therefore $(\eta \xi)^* = (\bar{\eta} \bar{\xi})$,
$\eta \xi = - \eta \xi$, $\bar{\eta} \bar{\xi} = - \bar{\xi} \bar{\eta}$,
and $\xi \xi = \bar{\xi} \bar{\xi} = 0$. ``Tilting'' of
indices: $\eta_{A} \xi^{A} = \eta^{B} \epsilon_{B A} \epsilon^{AC}
\xi_{C} = \eta^{B} ( - \delta^{C}_{B}) \xi_{C} = - \eta^{A} \xi_{A}$ and
the Schouten identity is
\begin{equation}
 \label{eq:schouten}
 \epsilon^{AB} \epsilon^{CD} + \epsilon^{AC} 
 \epsilon^{D B} + \epsilon^{AD} \epsilon^{BC} = 0\,.
\end{equation}
The vector of the Pauli matrices is defined by $\sigma^{\mu,\dot AB} =
(\mathbf{1}, \vec{\sigma})$ and $\bar{\sigma}^{\mu}_{A\dot{B}} =
(\mathbf{1}, - \vec{\sigma})$. We always distinguish the position of
the index:
\begin{equation}
 \sigma^1 =\; - \sigma_1 =  \begin{pmatrix}
 0 & 1 \\ 1 & 0 \end{pmatrix}, \;
 \sigma^2 = - \sigma_2 =  \begin{pmatrix}
 0 & -\ii \\ \ii & 0 \end{pmatrix}, \;
 \sigma^3 =\; - \sigma_3 =  \begin{pmatrix}
 1 & 0 \\ 0 & - 1 \end{pmatrix}\,.
\end{equation}
We conclude with some formulae for the spin tensors
that fix our conventions and makes the derivation of the spinor
representations more transparent. Hermiticity
$\sigma^{\mu,\dot A B} = \sigma^{\mu, B \dot{A}}$,
$\bar\sigma^{\mu}_{A \dot{B}} = \bar\sigma^{\mu}_{\dot{B} A}$;  
complex conjugation $\sigma^\mu_{\dot{A} B} =
(\bar\sigma^\mu_{A\dot{B}})^* = (\mathbf{1}, - \vec{\sigma}^*) = 
(\mathbf{1}, - \sigma^1, \sigma^2 , - \sigma^3)$, lowering
indices $\sigma^\mu_{\dot{A}B} = \sigma^{\mu,\dot{C}D}
\epsilon_{\dot{C}\dot{A}} \epsilon_{D B} = (\mathbf{1}, -\sigma^1,
\sigma^2, -\sigma^3)$ (using
$\sigma^\mu_{\dot{1}1} = \sigma^{\mu,\dot{2}2}$,
$\sigma^\mu_{\dot{2}2} = \sigma^{\mu,\dot{1}1}$,
$\sigma^\mu_{\dot{1}2} = -  \sigma^{\mu,\dot{2}1}$,
$\sigma^\mu_{\dot{2}1} = -  \sigma^{\mu,\dot{1}2}$) and finally
$\sigma_{\mu,\dot A B} =
 g_{\mu\nu} \sigma^\nu_{\dot A B} = (\mathbf{1}, \vec{\sigma}^*)$.

%%%%%%%%%%%%%%%%%%%%%%%%%%%%%%%%%%%%%%%%%%%%%%%%%%%%%%%%%%%%%%%%%%%%%%%%
\subsection{Decomposition of lightlike vectors}

Contraction of a Minkowskian 4-vector with the spin tensor results in
a spinor of rank two which is represented by a
$2\times2$ Hermitian matrix for real vectors
\begin{equation}
 K_{\dot AB} := k^{\mu} \sigma_{\mu, \dot AB} = k^\mu g_{\mu\nu}
 \sigma^\nu_{\dot AB} = k^0 \mathbf{1} - \vec{k} (-\vec{\sigma})^* = 
 \left(
 \begin{array}{cc} k^0 + k^3 & k^1 + \ii k^2 \\
 k^1 - \ii k^2 & k^0 - k^3 \end{array} \right)\,,
\end{equation}
which allows to write 4-vector products as
spinor products
\begin{equation}
 2 k \cdot p = k_{\mu} 2 g^{\mu \nu} p_{\nu} = k_{\mu} 
 \sigma^{\mu}_{\dot A B} \sigma^{\nu, \dot A B} 
 p_{\nu} =  K_{\dot A B} P^{\dot A B}\,.
\end{equation}
For lightlike momenta, the momentum spinor matrix can be written as a
tensor product 
\begin{equation}
  K_{\dot A B} = k_{\dot A} k_B , \qquad k_{\dot{A}} = (k_A)^*, \qquad
\text{with} \;\;
  k_A = 
    \begin{pmatrix}
    (p^1 - \ii p^2)/\sqrt{p^0-p^3}  \\
    \sqrt{p^0-p^3}
  \end{pmatrix},
\end{equation}
so that the spinor product is~\cite{BG,GaWu}
\begin{equation}
  \braket{pq} = (p_1 - \ii p_2) \frac{\sqrt{q_0-q_3}}{\sqrt{p_0-p_3}}
              - (q_1 - \ii q_2) \frac{\sqrt{p_0-p_3}}{\sqrt{q_0-q_3}}\,,
\end{equation}
and we find
\begin{equation}
  |\braket{pq}|^2 = 2p\cdot q\,.
\end{equation}

%%%%%%%%%%%%%%%%%%%%%%%%%%%%%%%%%%%%

\section{Constants, Expressions and Abbreviations}

Here we summarize expressions containing the noncommutativity and
polarization vectors:
\begin{subequations}
\begin{align}
  \left(p \theta \varepsilon_+(k)\right)
     &= \frac{\braket{p\phi g_+}^* \braket{pk}
                  + \braket{p\phi k} \braket{p
     g_+}^*}{2\sqrt2\braket{g_+k}^*}  
     \\ 
  \left(p \theta \varepsilon_- (k)\right)
     &= \frac{\braket{p\phi k}^* \braket{pg_-}
     + \braket{p\phi g_-} \braket{pk}^*}{2\sqrt2\braket{g_-k}}\,.
\end{align}
\end{subequations}
If the momentum is that of the photon, the gauge spinor cancels out, 
\begin{equation}
  \left(k \theta \varepsilon_+ (k)\right)
      =\; - \frac{1}{2\sqrt2} \braket{k\phi k}, \qquad
  \left(k \theta \varepsilon_- (k)\right)
      =\; - \frac{1}{2\sqrt2} \braket{k\phi k}^*.
\end{equation}
For the $(+,-)$ polarization of the photons we have 
\begin{equation}
  \left(\varepsilon_+(k_1) \theta \varepsilon_-(k_2)\right)
     = \frac{\Braket{g_+ k_2}^* \Braket{k_1 \phi g_-}
         + \Braket{k_1 g_-} \Braket{g_+ \phi k_2}^*}
          {2 \Braket{g_+ k_1}^* \Braket{g_- k_2}}\,.
\end{equation}
Translating vector to spinor expressions 
\begin{subequations}
\begin{align}
  (\varepsilon_1 \varepsilon_2) &= \frac{\Braket{g_+ k_2}^* \Braket{k_1
  g_-}}{\Braket{g_+k_1}^*\Braket{g_-k_2}} \\ 
  (k_1 \varepsilon_2) &= \frac{1}{\sqrt{2}} \frac{\Braket{k_1 k_2}^*
  \Braket{k_1 g_-}}{\Braket{g_-k_2}} \\  
  (k_2 \varepsilon_1) &= \frac{1}{\sqrt{2}} \frac{\Braket{k_2 g_+}^*
  \Braket{k_2 k_1}}{\Braket{g_+k_1}^*} \,.
\end{align}
\end{subequations}

%%%%%%%%%%%%%%%%%%%%%%%%%%%%%%%%%%%%%%%%%%%%%%%%%%%%%%%%%%%%%%%%%%%%%%%%
\section{The Feynman rules}
\label{app:feyn}

In the Feynman rules for helicity amplitudes, external fermions are
represented by
\begin{equation}
  \Psi_+ = \begin{pmatrix} k_A \\ 0 \end{pmatrix}, \quad
  \Psi_- = \begin{pmatrix} 0 \\ k^{\dot{A}} \end{pmatrix}, \quad
  \overline{\Psi}_+ = \left( 0 , k_{\dot{A}} \right), \quad
  \overline{\Psi}_- = \left( k^A , 0 \right) \,.
\end{equation}
The bispinor for an incoming antifermion is the same as the
outgoing fermion with the interchange $+ \leftrightarrow -$. (The
outgoing antifermion's bispinor is related to the incoming fermion's
bispinor in the same way).
Polarization vectors of incoming photons:
\begin{equation}
  \label{eps}
  \varepsilon_{+,\dot AB} (k)
     = \frac{\sqrt{2} g_{+,\dot A}k_{B}}{\braket{g_+k}^*}, \quad
  \varepsilon_{-,\dot AB} (k)
     = \frac{\sqrt{2} k_{\dot A}g_{-,B}}{\braket{g_-k}}\,.
\end{equation}
Fermion and $Z$ propagators in unitarity gauge:
\begin{equation}
  f\bar{f}: \;\;\; \frac{\ii}{k^2} \begin{pmatrix} 0 & 
  K_{A\dot{B}} \\ K^{\dot{A}B} & 0
  \end{pmatrix} , \qquad\quad Z_\mu Z_\nu: \;\;\; \frac{-2 \ii
  \epsilon_{\dot{A}\dot{C}} \epsilon_{BD}}{k^2 -
  M_Z^2 + \ii M_Z \Gamma_Z}\,.
\end{equation}
A $\gamma$ matrix coupling is translated to helicity amplitudes via 
\begin{equation}
  \gamma_\mu \left( C_L \frac{1-\gamma^5}{2} + C_R
  \frac{1+\gamma^5}{2} \right) \to \begin{pmatrix} 0 & 
  C_L \delta^{\dot{C}}_{\dot{B}} \delta^D_A \\
  C_R \epsilon^{\dot{A}\dot{C}} \epsilon^{BD} & 0
  \end{pmatrix}
\end{equation}
where $\dot C D$ are the spinor components of
the vector degree of freedom and $A, B$ the fermion spinor indices. 

The Feynman rules of the NCSM can be read
off from \cite{NCSM}. In order to simplify our formulae we introduce
partial contractions 
\begin{equation}
  (k\theta)^\nu = k_\mu \theta^{\mu\nu}, \qquad
  (\theta k)^\mu = \theta^{\mu\nu} k_\nu, \qquad
  (k\theta)^\mu = - (\theta k)^\mu\,.
\end{equation}
Vertices with \emph{all momenta outgoing}
\begin{subequations}
\begin{align}
  \parbox{35mm}{\includegraphics[width=35mm]{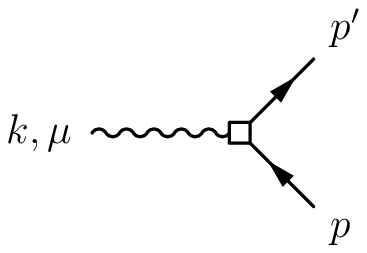}}
    &= \ii g \left(\gamma_\mu + \Gamma_\mu(k,p)\right) \\
  \parbox{35mm}{\includegraphics[width=35mm]{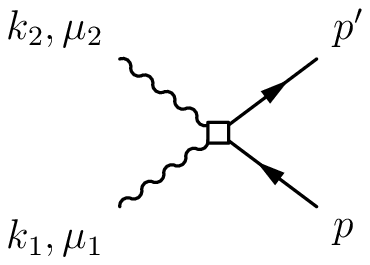}}
    &= \ii g^2 H_{\mu_1\mu_2} (k_1,k_2)
\end{align}
\end{subequations}
with
\begin{subequations}
\begin{equation}
  \Gamma_\mu(k,p) =
    - \frac{\ii}{2} \left[
         (k\theta p) \gamma_{\mu}
        - (k\theta)_\mu \fmslash{p}
        - (\theta p)_\mu \fmslash{k}
       \right] = - \Gamma_\mu(p,k)
\end{equation}
\begin{align}
  H_{\mu_1\mu_2} (k_1,k_2) &=
    - \frac{\ii}{2} \left[
       \theta_{\mu_1\mu_2} (\fmslash{k}_1-\fmslash{k}_2)
        + ((k_1-k_2) \theta)_{\mu_1} \gamma_{\mu_2}
        - ((k_1-k_2) \theta)_{\mu_2} \gamma_{\mu_1}
       \right] \notag \\
    &= H_{\mu_2\mu_1} (k_2,k_1)\,.
\end{align}
\end{subequations}
And
\begin{equation}
  \parbox{45mm}{\includegraphics[width=45mm]{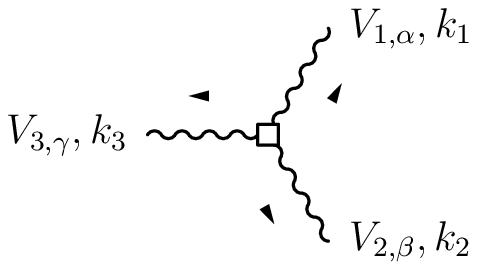}}
     =: \ii \mathcal{V}_{V_1V_2V_3}
\end{equation}
with
\begin{multline}
  V_{\alpha\beta\gamma}(k_1,k_2,k_3)
    =   \theta_{\alpha\beta}
          \left[ (k_1 k_3) k_{2,\gamma} - (k_2 k_3) k_{1,\gamma} \right]
      + \left( k_1 \theta k_2 \right)
          \left[ k_{3,\alpha} \eta_{\beta\gamma}
                   - \eta_{\alpha\gamma} k_{3,\beta} \right] \\
      + \Bigl[   (k_1\theta)_\alpha
                   \left[ k_{2,\gamma} k_{3,\beta} - (k_2 k_3) g_{\beta\gamma} \right]
               - (\alpha \leftrightarrow \beta)
               - (\alpha \leftrightarrow \gamma)
        \Bigr] \\
  + \text{cyclical permutations of}\;
       \bigl\{ (\alpha,k_1) , (\beta,k_2), (\gamma,k_3) \, \bigr\}\,,
\end{multline}
the $\gamma_\alpha(k_1)\gamma_\beta(k_2)Z_\gamma(p)$ and
$\gamma_\alpha(k_1)\gamma_\beta(k_2)\gamma_\gamma(k_3)$ vertices are
given by   
\begin{align}
  \ii \mathcal{V}_{Z\gamma\gamma} =&\; + 2 e \sin(2\theta_W) \,
  K_{Z\gamma\gamma} \cdot V_{\alpha\beta\gamma}(k_1,k_2,p) \\ 
  \ii \mathcal{V}_{\gamma\gamma\gamma} =&\; - 2 e \sin(2\theta_W) \,
  K_{\gamma\gamma\gamma} \cdot V_{\alpha\beta\gamma}(k_1,k_2,k_3)
\end{align}
The coupling constants are related to the electroweak coupling constants by
\begin{subequations}
\begin{align}
  K_{\gamma\gamma\gamma} &=\; \frac{1}{2} g g' \left( \kappa_1 + 3
  \kappa_2 \right), \\
  K_{Z\gamma\gamma} &=\; \frac{1}{2} \Bigl[ g^{\prime\, 2} \kappa_1 +
  \left( g^{\prime\, 2} - 2 g^2 \right) \kappa_2 \Bigr],
\end{align}
\end{subequations}
where $\kappa_{1/2}$ are the parameters defined in \cite{TGC}.

%%%%%%%%%%%%%%%%%%%%%%%%%%%%%%%%%%%%%%%%%%%%%%%%%%%%%%%%%%%%%%%%%%%%%%%%
\section{Standard Model Helicity Amplitudes}
\label{app:helamp}

Amplitudes with like fermion helicities are zero
\begin{equation}
    A (\sigma_1,\sigma_2,+,+) = A (\sigma_1,\sigma_2,-,-) = 0
\end{equation}
and the SM amplitudes are (cf.~also \cite{GaWu})  
\begin{subequations}
\begin{align}
  A_{\text{SM}}(+,-,+,-) &=\;  - 2 \ii e^2 Q_f^2
  \frac{\Braket{p_2 k_1} \Braket{p_1 k_2}^*}{\Braket{p_1 k_2}
  \Braket{p_1 k_1}^*} \\ 
  A_{\text{SM}}(+,-,-,+) &=\;  - 2 \ii e^2 Q_f^2
  \frac{\Braket{p_1 k_1} \Braket{p_2 k_2}^*}{\Braket{p_1 k_2}
  \Braket{p_1 k_1}^*}\,.
\end{align}
\end{subequations}
The combinations with the reversed $(-,+)$ photon
polarizations are determined from the $(+,-)$ combination
by interchanging $k_1$ and $k_2$. 

%%%%%%%%%%%%%%%%%%%%%%%%%%%%%%%%%%%%%%%%%%%%%%%%%%%%%%%%%%%%%%%%%%%%%%%%
%%% References
%%%%%%%%%%%%%%%%%%%%%%%%%%%%%%%%%%%%%%%%%%%%%%%%%%%%%%%%%%%%%%%%%%%%%%%%


\baselineskip15pt
\begin{thebibliography}{99}

\bibitem{SeiWit}
  N.~Seiberg, E.~Witten, 
  %``String Theory and Noncommutative Geometry'' 
  J. High Energy Phys.{} \textbf{JHEP09} (1999) 032
  [arXiv:hep-th/9908142].
%%CITATION = hep-th/9908142;%

\bibitem{UVIR}
  S.~Minwalla, M. Van Raamsdonk, N.~Seiberg,
  %``Noncommutative Perturbative Dynamics''
  J. High Energy Phys.{} \textbf{JHEP02} (2000) 020
  [arXiv:hep-th/9912072].
  %%CITATION = hep-th/9912072;%

\bibitem{Unitarity}
  J.~Gomis, T.~Mehen,
  %%``SPACE-TIME NONCOMMUTATIVE FIELD THEORIES AND UNITARITY.''
  Nucl.{} Phys.{} \textbf{B591} (2000) 265
  [arXiv:hep-th/\allowbreak0005129];
  %%CITATION = hep-th/0005129;%%
  D.~Bahns, S.~Doplicher, K.~Fredenhagen, G.~Piacitelli,
  %%``ON THE UNITARITY PROBLEM IN SPACE-TIME NONCOMMUTATIVE THEORIES.''
  Phys.{} Lett.{} \textbf{B533} (2002) 178
  [arXiv:hep-th/0201222];
  %%CITATION = hep-th/0201222;%
  Yi Liao, K.~Sibold,
  %%''TIME ORDERED PERTURBATION THEORY ON NC SPACE-TIME. 2. UNITARITY.''
  Eur.{} Phys.{} J.{} \textbf{C25} (2002) 479
  [arXiv:hep-th/0206011];
  %%CITATION = hep-th/0206011;%%
  T.~Ohl, R. R\"uckl, J.~Zeiner,
  %%''UNITARITY OF TIME - LIKE NONCOMMUTATIVE GAUGE THEORIES...''
  Nucl.{} Phys.{} \textbf{B676} (2004) 229
  [arXiv:hep-th/0309021].
  %%CITATION = hep-th/0309021;%%

\bibitem{Hinchliffe}
  I.~Hinchliffe, N.~Kersting,
  %%''CP VIOLATION FROM NONCOMMUTATIVE GEOMETRY.''
  Phys.{} Rev.{} \textbf{D64} (2001) 116007
  [arXiv:hep-ph/0104137].
  %%CITATION = hep-ph/0104137;%%

\bibitem{HewettRizzo}
  J.~L.~Hewett, F~.J.~Petriello, T.~G.~Rizzo,
  %%``SIGNALS FOR NONCOMMUTATIVE INTERACTIONS AT LINEAR COLLIDERS''
  Phys.{} Rev.{} \textbf{D64} (2001) 075012
  [arXiv:hep-ph/0010354];
  %%CITATION = hep-ph/0010354;%
  S.~Godfrey, M.~A.~Doncheski,
  %%``SIGNALS FOR NONCOMMUTATIVE QED IN E GAMMA AND GAMMA GAMMA COLLISIONS.'' 
  Phys.{} Rev.{} \textbf{D65} (2002) 015005
  [arXiv:hep-ph/0108268];
  %%CITATION = hep-ph/0108268
  S.~Godfrey, M.~A.~Doncheski,
  %%''PHENOMENOLOGY OF NONCOMMUTATIVE FIELD THEORIES.''
  Talk at the \textit{10th International Conference on Supersymmetry and
  Unification of Fundamental Interactions (SUSY02), Hamburg, Germany, 17-23 Jun 2002},
  [arXiv:hep-ph/0211247].
  %%CITATION = hep-th/0211247;%%

\bibitem{Wess:pr}
  J.~Wess,
  %``Non-Abelian Gauge Theories On Non-Commutative Spaces,''
  Commun.{} Math.{} Phys.{}  \textbf{219} (2001) 247.
%%CITATION = CMPHA,219,247;%%

\bibitem{Armoni}
  A.~Armoni,
  %%% COMMENTS ON PERTURBATIVE DYNAMICS OF NONCOMMUTATIVE YANG-MILLS THEORY. 
  Nucl.{} Phys.{} \textbf{B593} (2001) 229
  [arXiv:hep-th/0005208].
  %% CITATION = hep-th/0005208;%

\bibitem{Hayakawa:1999yt}
  M.~Hayakawa,
  %``Perturbative analysis on infrared aspects of noncommutative QED on  R**4,''
  Phys.{} Lett.{} \textbf{B478} (2000) 394
  [arXiv:hep-th/9912094].
  %%CITATION = HEP-TH 9912094;%%

\bibitem{Sheikh-Jabbari}
  L.~Bonora, M.~Schnabl, M.~M.~Sheikh-Jabbari, A.~Tomasiello, 
  %%``NONCOMMUTATIVE SO(N) AND SP(N) GAUGE THEORIES''
  Nucl.{} Phys.{} \textbf{B589} (2000) 461
  [arXiv:hep-th/0006091];
  %%CITATION = hep-th/0006091;%
  M.~Chaichian, P.~Presnajder, M.~M.~Sheikh-Jabbari, A.~Tureanu,
  %%% NONCOMMUTATIVE GAUGE FIELD THEORIES: A NO GO THEOREM.
  Phys.{} Lett.{} \textbf{B526} (2002) 132
  [arXiv:hep-th/0107037],
  %% CITATION = hep-th/0107037;%
  %%% M.~Chaichian, P.~Presnajder, M.~M.~Sheikh-Jabbari, A.~Tureanu
  %%% NONCOMMUTATIVE STANDARD MODEL: MODEL BUILDING.
  Eur.{} Phys.{} J.{} \textbf{C29} (2003) 413
  [arXiv:hep-th/0107055].
  %% CITATION = hep-th/0107055;%

\bibitem{NCSM}
  X.~Calmet, B.~Jur{\v c}o, P.~Schupp, J.~Wess, M.~Wohlgenannt,
  %``The Non-Commutative Standard Model''
  Eur.{} Phys.{} J.{} \textbf{C23} (2002) 363
  [arXiv:hep-ph/0111115];
  %% CITATION = hep-ph/0111115;%
  X.~Calmet, M.~Wohlgenannt,
  %%''EFFECTIVE FIELD THEORIES ON NONCOMMUTATIVE SPACE-TIME.''
  Phys.{} Rev.{} \textbf{D68} (2003) 025016
  [arXiv:hep-ph/0305027].
  %%CITATION = hep-ph/0305027;%%

\bibitem{thojrprog}
  T.~Ohl, J.~Reuter, in preparation. 

\bibitem{Raffelt}
  P.~Schupp, J.~Trampeti\'c, J.~Wess, G.~Raffelt, 
  %%``THE PHOTON NEUTRINO INTERACTION IN NONCOMMUTATIVE GAUGE.....''
  Eur.{} Phys.{} J.{} \textbf{C} (in print)
  [arXiv:hep-ph/0212292].
  %%CITATION = hep-ph/0212292;%%

\bibitem{raredec}
  J.~Trampeti\'c,
  %%``RARE AND FORBIDDEN DECAYS.''  
  Acta\ Phys.{} Polon.{} \textbf{B33} (2002) 4317
  [arXiv:hep-ph/\allowbreak0212309];
  %%CITATION = hep-ph/0212309;%
  P.~Minkowski, P.~Schupp and J.~Trampeti\'c,
  %``Non-commutative '*-charge radius' and '*-dipole moment' of the  neutrino,''
  [arXiv:hep-th/0302175];
  %%CITATION = HEP-TH 0302175;%%
  P.~Schupp, J.~Trampeti\'c,
  %%''THE NONCOMMUTATIVE STANDARD MODEL AND FORBIDDEN DECAYS.''  
  [arXiv:hep-ph/0405163].
  %%CITATION = hep-ph/0405163;%  

\bibitem{TGC}
  N.~G.~Deshpande, X.-G.~He,
  %% TRIPLE NEUTRAL GAUGE BOSON COUPLINGS IN NONCOMMUTATIVE STANDARD MODEL. 
  Phys.{} Lett.{} \textbf{B533} (2002) 116
  [arXiv:hep-ph/0112320];
  %% CITATION = hep-ph/0112320;%
  W.~Behr, N.~G.~Deshpande, G.~Duplan{\v c}i\'c, P.~Schupp,
  J.~Trampeti\'c, and J.~Wess,
  % Triple gauge couplings in NCSM
  Eur.{} Phys.{} J.{} \textbf{C 29} (2003) 441
  [arXiv:hep-ph:0202121];
  %% CITATION = hep-ph/0202121;%
  G.~Duplan{\v c}i\'c, P.~Schupp, J.~Trampeti\'c, 
  % A comment on the above...
  Eur.{} Phys.{} J.{} \textbf{C 32} (2003) 141
  [arXiv:hep-ph/0309138].
  %% CITATION = hep-ph/0309138;%

\bibitem{anatho}
  A.~Alboteanu, T.~Ohl,
  %%% NCSM at a Hadron Collider
  in preparation.

\bibitem{ncbounds}
  C.~E.~Carlson, C.~D.~Carone, R.~F.~Lebed,
  %%% BOUNDING NONCOMMUTATIVE QCD.
  Phys.{} Lett.{} \textbf{B518} (2001) 201
  [arXiv:hep-ph/0107291];
  %% CITATION = hep-ph/0107291;%
  C.~E.~Carlson, C.~D.~Carone, R.~F.~Lebed,
  %%% SUPERSYMMETRIC NONCOMMUTATIVE QED AND LORENTZ VIOLATION.
  Phys.{} Lett.{} \textbf{B549} (2002) 337
  [arXiv:hep-ph/0209077];
  %% CITATION = hep-ph/0209077;%
  P.~Castorina, A.~Iorio, D.~Zappala,
  %%% NONCOMMUTATIVE SYNCHROTRON.
  Phys.{} Rev.{} \textbf{D69} (2004) 065008
  [arXiv:hep-th/0212238];
  %% CITATION = hep-th/0212238;%
  P.~Castorina, D.~Zappala,
  %%% NONCOMMUTATIVE ELECTRODYNAMICS AND ULTRAHIGH-ENERGY GAMMA-RAYS.
  Europhys.{} Lett.{} \textbf{64} (2003) 641
  [arXiv:hep-ph/0310116];
  %% CITATION = hep-ph/0310116;%
  X.~Calmet,
  %%% WHAT ARE THE BOUNDS ON SPACE-TIME NONCOMMUTATIVITY?
  CALT-68-2473, [arXiv:hep-ph/0401097].
  %% CITATION = hep-ph/0401097;%

\bibitem{PDG}
  K. Hagiwara et al. (Particle Data Group), 
  Phys. Rev. D \textbf{66} (2002) 010001.

\bibitem{TDR}
  B. Badelek et al., 
  \textit{TESLA: The Superconducting Electron Positron Linear Collider With An
  Integrated X-Ray Laser Laboratory. Technical Design Report. Part 6. 
  Appendices. Chapter 1. Photon Collider At Tesla.},    
  [arXiv:hep-ex/0108012].
  %%CITATION = hep-ex/0108012;%%

\bibitem{Ginzburg/Telnov/etal:gamma-collider}
  I.~F.~Ginzburg, G.~L.~Kotkin, V.~G.~Serbo, V.~I.~Telnov,
  JETP Lett.{} \textbf{34}, 491 (1981);
%%CITATION = JTPLA,34,491;%%
  Nucl.{} Instrum.{} Meth.{}  \textbf{205} (1983) 47;
%%CITATION = NUIMA,205,47;%%
  I.~F.~Ginzburg, G.~L.~Kotkin, S.~L.~Panfil, V.~G.~Serbo, V.~I.~Telnov,
  Nucl.{} Instrum.{} Meth.{} A \textbf{219} (1984) 5.
%%CITATION = NUIMA,A219,5;%%

\bibitem{BG}
  F.~A.~Berends, W.~Giele, 
  %%THE SIX GLUON PROCESS AS AN EXAMPLE OF WEYL-VAN DER WAERDEN SPINOR
  %%CALCULUS.
  Nucl.{} Phys.{} \textbf{B294} (1987) 700;
  S.~Dittmaier,
  %%%WEYL-VAN DER WAERDEN FORMALISM FOR HELICITY AMPLITUDES OF MASSIVE
  %%%PARTICLES. 
  Phys.{} Rev.{} \textbf{D59} (1999) 016007
  [arXiv:hep-ph/9805445].
  %% CITATION = hep-ph/9805445;%

\bibitem{GaWu}
  R.~Gastmans, T.~T.~Wu, 
  \textit{The Ubiquitous Photon -- Helicity Method for QED and QCD} 
  (Oxford UP, 1990).

\bibitem{Penrose:book}
R.~Penrose, W.~Rindler, \textit{Spinor Calculus and Twistor
  Geometry} (Addi\-son-Wesley, 1995), pp.~147ff, 312ff.

\bibitem{Omega}
  T. Ohl, \textit{O'Mega: An Optimizing Matrix Element Generator},
  in \textit{Proceedings
  of 7th International Workshop on Advanced Computing and Analysis Techniques
  in Physics Research (ACAT 2000)} (Fermilab, Batavia, Il, 2000)
  [arXiv:hep-ph/0011243];
%%CITATION = HEP-PH 0011243;%%
  M.~Moretti, T.~Ohl, J.~Reuter,
  %``O'Mega: An optimizing matrix element generator,''
  [arXiv:hep-ph/0102195].
%%CITATION = HEP-PH 0102195;%%

\bibitem{WHIZARD}
  \url{http://whizard.event-generator.org}; W.~Kilian, T.~Ohl and
  J.~Reuter,
  %``WHIZARD: Simulating Multi-Particle Processes at LHC and ILC,''
  arXiv:0708.4233 [hep-ph].
  %%CITATION = ARXIV:0708.4233;%%

\bibitem{Circe}
  T.~Ohl, 
  %%``CIRCE VERSION 1.0: BEAM SPECTRA FOR SIMULATING LINEAR COLLIDER PHYSICS.''
  Comput.{} Phys.{} Commun.{} \textbf{101} (1997) 269
  [arXiv:hep-ph/\allowbreak9607454];
  %%CITATION = hep-ph/9607454;%%
  T.~Ohl, {\tt Circe 2.0} Beam Spectra for Simulating Linear Collider
  and Photon Collider Physics, WUE-ITP-2002-006.

\bibitem{Telnov}
  V. Telnov, private communication.

\end{thebibliography}
\end{document}